\documentclass[preprint,12pt,3p]{elsarticle}

\usepackage{amssymb}

\journal{Elsevier}
\usepackage{color}
\usepackage{amssymb}
\usepackage{array}
\usepackage{booktabs}
\usepackage{multirow}
\usepackage{subfigure}
\usepackage{graphicx}
\usepackage{dcolumn}
\usepackage{bm}

\begin{document}

\begin{frontmatter}

\title{Correlation between photoemissive and morphological properties of KBr thin film photocathodes}

\author[add1]{R. Rai}
\author[add2]{Triloki}
\author[add1]{B. K.~Singh\corref{cor}}
\cortext[cor]{Corresponding author}
\ead{bksingh@bhu.ac.in}
\author[add1]{N. F.A. Jammal}
\address[add1]{Department of Physics, Institute of Science, Banaras Hindu University, Varanasi-221005, India.}
\address[add2]{Solid State and Structural Chemistry Unit, Indian Institute of Science, Bengaluru-560012, India.}

\begin{abstract}
In the present work, the morphological properties of KBr photocathodes are correlated with their photoemissive behavior using a combination of analyzing techniques including SEM, TEM and AFM. From morphological studies, it is observed that KBr films have granular characteristics with varied average grain size and grain density. Structure and orientation of individual grains have been investigated by selected area electron diffraction technique and found to be the crystalline in nature with a face centered cubic structure. It is evident from the AFM analysis that the root mean square roughness and maximum area peak height have been decreased with the deposition of more KBr layers. The photoemission studies reveal that the resultant photocurrent is enhanced with increasing film thickness and it is directly related to the surface area coverage and grain density. 

 \end{abstract}


\end{frontmatter}


\section{Introduction}
\label{sec1}

Photocathodes will always be an essential part of high energy and astrophysics detector systems for converting an incoming photons into primary photoelectrons. The most critical task is to choose a photocathode material with  high quantum yield, a single photon sensitivity, good stability and fast response time. From last two decades, alkali halides are emerging as an efficient photo converter in extreme ultraviolet (EUV, 10 nm \textless $\lambda$ \textless 100 nm) and vacuum ultraviolet (VUV, 100 nm \textless $\lambda$ \textless 200 nm) wavelength ranges~\cite{Tril_Str}. These photocathode materials combine high VUV quantum yields with the ability to withstand a moderate amount of exposure to humid air and it makes them a feasible choice for UV and soft X-ray instrumentations~\cite{OHW}. The relatively high photoelectric work function of these materials results in cutoff between 150 to 300 nm; reduces the sensitivity at longer spectral ranges and lowers the dark count rates by minimizing the  response to the visible band. At the shorter wavelengths (typically $<$160 nm), potassium bromide (KBr) photocathodes are one of the promising candidates to maximize the quantum detection efficiency of MCP based detectors ~\cite{Leesa,Os,Yoshioka} and employed in many astrophysics experiments such as;  the UV spectroscope PHEBUS on Board of ESA/JAXAs BepiColombo Mercury exploration mission~\cite{PHEBUS}, SUMER on the Solar and Heliospheric Observatory (SOHO)~\cite{SUMER}, the Far Ultraviolet Spectroscopic Explorer (FUSE)~\cite{over}, the Cassini Ultraviolet Imaging Spectrograph~\cite{UVIS}, XUV on the NOZOMI~\cite{NOZOMI}, etc. These photocathodes must be protected from the ambient atmosphere, but can be operated stably in the space. The solar blind response of KBr photocathodes improve the signal quality by rejecting the background near UV wavelengths~\cite{Oswald,ASTr} and more over, these photocathodes also serve as a protecting layer for the visible sensitive detectors~\cite{A. Breskin1996}. Although, enormous studies have been performed on the photoemission measurement of KBr thin film, in order to make them a potential alternative for detector applications~\cite{Tremsin,Weidong}, but very less works have been done in terms of their surface characterization. In connection to this, the morphological investigations are an important aspect to optimize the photocathode fabrication process. 

The thin films grown from physical vapor-deposition processes such as sputtering, thermal evaporation or chemical vapor deposition, usually have a columnar morphology~\cite{AMMERS}. The study of two basic factors: surface roughness, $\sigma$ and a characteristic lateral correlation length or grain size, $\xi$, provided a clear understanding regarding the film morphology~\cite{Vasco}. It is assumed that $\sigma$ and $\xi$ change with the film thickness, $d$, as $\sigma \sim d/g^{\beta}$ and $\xi \sim d/g^{p}$, respectively, here $g$ is the deposition rate (thickness of the layer deposited per second)~\cite{Smilauer}. The exponents $\beta$ and $p$ depend on the growth mechanisms operating during the deposition~\cite{Barabási,Meakin}. The parameters $\sigma$ and $\xi$ can be directly obtained by acquiring a topological images. Therefore, in the present work, we have employed different characterization tools in order to illustrate the surface morphology of the freshly deposited films and obtained results are correlated with their photoemissive behavior.

\section{Experimental Details}

Thin films were deposited in the vacuum chamber by evaporation of 5N pure KBr powder through resistive vapor evaporation technique; tantalum ($Ta$) boat was used as a resistive source and cleaned substrates were placed face down, vertical direction at 25 cm distance above the boat. The chamber was evacuated with a turbo molecular pump system and the base pressure and pressure during deposition were $5.6\times 10^{-7}$ Torr and $\sim10^{-6}$ Torr, respectively. Prior to deposition, composition of residual gases and water vapor was analyzed by a quadrapole mass spectrometer (SRS 200). The chamber was dominated by $N_{2}$ (36.5\%) and traces of other gases were found in relatively small amount $(\textless 37.10\%)$. At this stage, the deposition started by heating KBr powder up to a melting point. Film thickness and deposition rate ($\leq 3nm/s$) were controlled through $Qz$ crystal based thickness/rate monitor. To avoid the contact with the atmospheric air, the deposition chamber was flushed with dry $N_{2}$  gas and KBr films were extracted into a vacuum desiccator for its transportation to morphological and photoemission measurements.

Atomic Force Microscopy (AFM), Scanning Electron Microscopy (SEM) and Transmission electron Microscopy (TEM) techniques were used for the morphological investigations of KBr thin film photocathodes. AFM ND-MDT solver-NEXT~\cite{AFM}, coupled with the PX Ultra controller was employed to acquire the surface topology of films at the atomic scale. Data acquisition and off line analysis  were performed by NOVA PX data processing software. SEM images of KBr film were recorded by FEI Quanta $200$ at 10-15 kV accelerating voltage with secondary electron detector in $5\times10^{-4}$~Torr vacuum environment. As KBr is an insulator therefore before installation into the SEM chamber, the electrical contact has been made between film and stubs using silver paste in order to minimize the possibility of sample discharging and burning during measurements.  TEM micrographs were taken by TECNAI-20$G^{2}$, operated at 200 kV voltage in the bright field diffraction and imaging modes with single tilt sample stage
\begin{figure}[!ht]
 \begin{center}
\includegraphics[scale=.48]{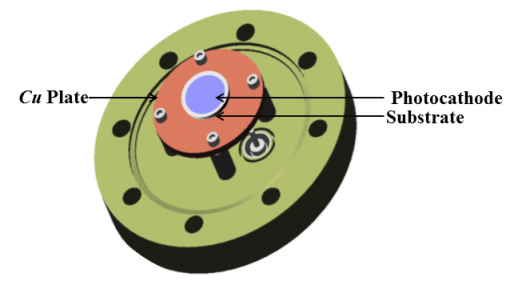} 
 \caption{The schematic representation of sample holder.}
     \label{XRD}
    \end{center}
  \end{figure}

30 $W$ $MgF_{2}$ windowed deuterium lamp emits UV photons, a monochromator (234/302, McPherson) was utilized to allow a selection of wavelength for photocurrent measurements. The collimation optics, together with reflecting mirrors, were exploited to focalize the light beam. An aluminum coated pick of mirror was placed at $45^{o}$ in the photocathode housing to split the UV light into two parts; 50\% photons impinge on photocathode and 50\% on scintillated PMT, which was used to monitor the stability of the lamp. Figure 1 illustrates the schematic representation of sample holder, used for mounting the KBr film in the photocurrent measurement set-up. A $Cu$ plate with a cavity in the middle of it, is attached to provide the electrical contact with the substrate. The photocathode is placed between the cavity and high voltage applied to it.  The entire set-up operates under vacuum and the base pressure of the system during the measurement was $1.6\times 10^{-5}$ Torr.

\newcommand{\head}[1]{\textnormal{\textbf{#1}}}
\newcommand{\normal}[1]{\multicolumn{1}{l}{#1}}
\section{Results and Discussion} 

\subsection{Morphological studies}

 The morphology of a growing films is determined by the growth competition between adjacent grains. For morphological measurements, KBr films were evaporated on different substrates; for AFM on stainless steel (S. S.) disc of 1 cm diameter, for SEM on S. S. stubs and for TEM it has grown directly on formvar coated copper grid. To acquire morphology with varying thickness, entire film surface was scanned; however, in this article we have only shown the images of few selected areas. All these reported micropatterns were recorded  in relative humidity (RH = 46 $\pm$~5$\%$) and at 24 $\pm$ 3$~^{o}C$ temperature.

\subsubsection{AFM analysis}
\begin{figure}[!ht]
 \begin{center}
\includegraphics[scale=2.9]{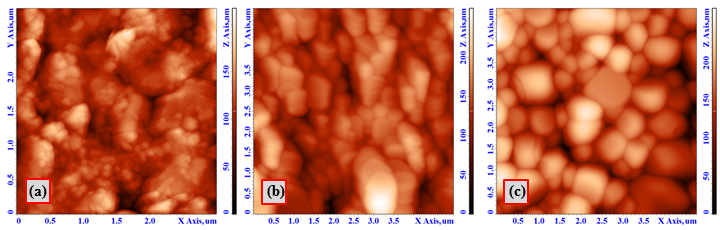} 
 \caption{ AFM images of (a) 200, (b) 300, and (c) 500 nm KBr films.}
     \label{XRD}
    \end{center}
  \end{figure} 

 The $2D$ AFM  patterns of $5\times 5 ~\mu m^{2}$ regions, are presented in Figure 2(a)-(c). It is evident from the images that thermally deposited KBr films mainly have a granular characteristic. The grain size, $\xi$ , is related to the lateral size of the granular structure and with the increasing film thickness; these grains are  growing at the expense of smaller ones. The average value of grain sizes, $\xi_{av}$ are 73, 120, 376 nm for 200, 300, 500 nm films respectively. The growth in grain size is observed due to the continuous bombardment of energetic particles during the film formation. This bombardment promotes the alteration of lateral length by the enhancement of nucleation and interdiffusion of grain boundary. The film surface is then modified through elimination of interfacial voids and porosity with creation of more isotropic grain morphology; results in extension of grains~\cite{ming}. 

The average roughness, $\sigma_{avg}$ represents the mean value of the surface height ($\bar{Z}$) relative to the center plane and it is found to be about  32, 29 and 22 nm for 200, 300, 500 nm films respectively. The root-mean-squared roughness, $\sigma_{rms}$ is determined by the standard deviation in $Z$ height and its value obtained by morphology of $5\mu m\times 5\mu m$ cross over image area is  41, 37, 28 nm for 200, 300, 500 nm respectively. AFM analysis depicts  that the surface roughness decreases  with increasing film thickness. The considerable change in a roughness is evident due to elimination of surface defects with elevation of film thickness. The maximum area peak height, $S_{p}$ for 200 nm film is 140 and for 300 and 500 nm film it is  110 and 67 nm respectively. The possible mechanism involves in the reduction of $S_{p}$ can be explained by Van der Drift model named as a ``principle of evolutionary selection"~\cite{Drift}. This model describe the competition between evaporating crystals during the film formation. As growth proceeds, more and more crystals are overgrown by adjacent crystals and the number of crystals extending to the surface is decreased which reduce the peak area with increasing thickness~\cite{Wild}.

\subsubsection{SEM analysis}

\begin{figure}[!ht]
 \begin{center}
\includegraphics[scale=0.71]{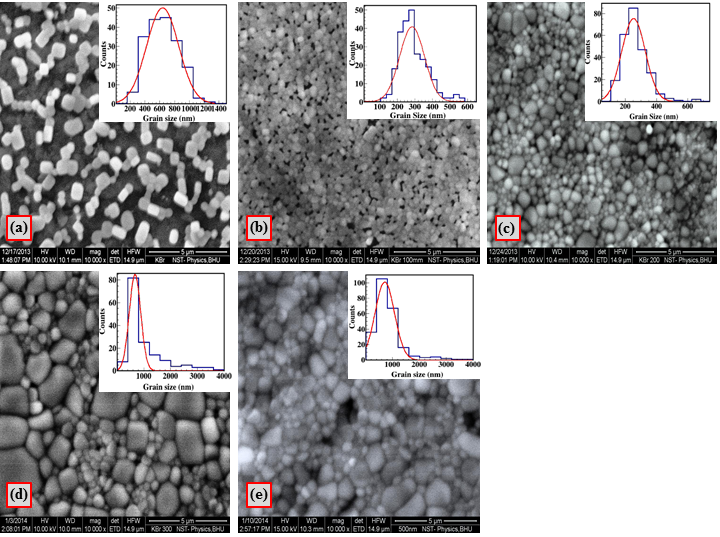} 
 \caption{SEM and grain size distribution images (inset) of (a) 50, (b) 100, (c) 200, (d) 300, (e) 500 nm KBr photocathode.}
     \label{XRD}
    \end{center}
  \end{figure}

 The surface morphology and the grain size distribution (inset) obtained by SEM is shown in Figures 3 (a)-(e). The value of $\xi_{av}$ was estimated from fitting the grain size distributions by a Gaussian function. As can be seen in Figure 3(a), 50 nm  KBr  layer  is  found to  be  fairly uniform with discontinuous square and rectangular structure. It may be because KBr crystals are face centered cubic and films also have the same structure. The sizes of the rectangular structures are varied from 203 to 802 nm and the surface area coverage is only around 40\%. The root mean square and average size are about 145 and 524 nm respectively.  However, 100 nm thick film exhibits a morphology of interconnected  grains  (Figure~3(b)) and their sizes, $\xi$ are  varying from 228 to 555 nm. The surface area coverage has also improved from 40\% to 70\% and $\xi_{av}$ is about  284 nm in diameter. As shown in Figure 3(c) that in the case of 200 nm film, again grain like morphology is observed with sizes varying from 216 to 1260 nm and grain density is increased as compared to 50 and 100 nm films. The  value of $\xi_{av}$ is found to be about 452 nm. For 300 and 500 nm films, surface area  is manifested by bigger grain with almost full substrate coverage. For 300 nm film grain sizes are varying from 144 to 2448 nm and the root mean square and $\xi_{av}$ are found to be about 375 and 596 nm respectively. However, for 500 nm film, grains are varying between 378 to 1697 nm and $\xi_{av}$ is 568 nm in diameter. It is evident from the images that grain sizes as well as grain density are increased with increasing film thickness. Only for 50 nm film, we obtained larger grain size than the thicker films, this is because of the hygroscopic nature of KBr and it is possible that during the transfer from deposition unit to SEM set-up sample may come into contact with humid air  which results in extension in the grain size due coalescence~\cite{Rai}. The rectangular or cubic shape may be observed due to the recrystallization results from the diffusion of adjacent grains. Average grain size  and surface area coverage of 300 and 500 nm films are approximately same which indicates that after certain thickness these values saturate. Although grain density of 500 nm film is higher than that of 300 nm film.

\subsubsection{TEM analysis}
 
 \begin{figure}[!ht]
 \begin{center}
\includegraphics[scale=0.69]{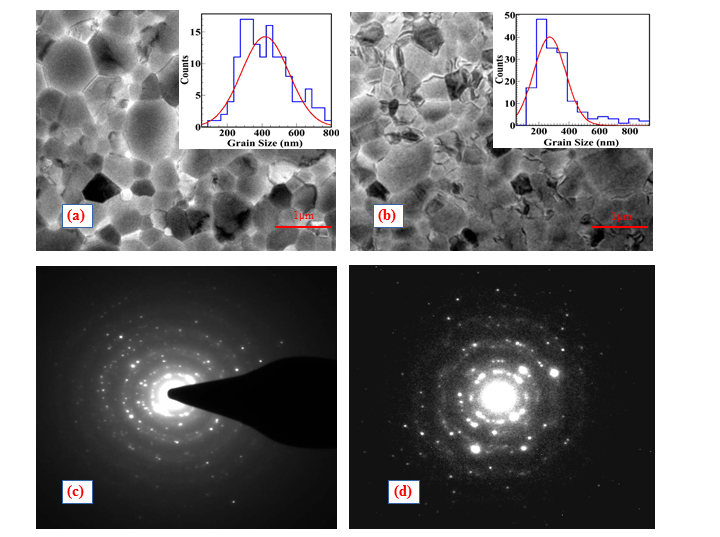} 
 \caption{ TEM and grain size distribution (inset) images of (a) 300 (b) 500 nm KBr film with the diffraction patterns (c) and (d).}
     \label{XRD}
    \end{center}
  \end{figure}
   
 TEM technique was employed to extract the information about the structure and the real size distribution of the KBr photocathodes and obtained results are shown in Figure 4. It can be evident from TEM observations that KBr films have a granular characteristics, but with a different grain size distributions. The estimated grain size distribution from 300 and 500 nm films are presented in the inset of Figure 4(a) and (b). The average grain size evaluated from Gaussian fitting of the grain size distribution comes out to be 413 nm for 300 nm film. Similarly, the $\xi_{av}$ determined by the fitting of size distribution of 500 nm film is about 309 nm. In order to study the microstructure and phase of the granular system, the selected area electron diffraction (SAED) pattern was recorded by focusing an electron beam onto a particular grain of interest. The obtained SAED patterns reflect that KBr films have a poly crystalline structure and orientation. In this study, almost all grains are analyzed and found that they are definitive a fcc crystallites with the precisely matching the lattice constant for the bulk KBr. The lattice constant of KBr films are manually calculated for 150 mm camera length (camera constant = 4.8466 cm\AA) and it found to be about 6.3284 \AA~ for 300 nm film and 6.4281 \AA~for 500 nm film. These values are slightly smaller than the ASTM card (pdf no.:730381) data ($a$ = 6.5847 \AA) reported for strain free crystal. This deviation in the lattice constant value from perfect lattice is attributed to the existence of lattice distortion due to the finite length of coherent scattering domain and the microstrain.

It is evident from AFM, SEM and TEM images that KBr films have faceted granular morphology with varying size distribution. Although, the sizes observed from these morphological tools are different, but follow the similar trend $i.e.$ increased with increasing film thickness. This difference is appeared due to the dissimilarity of the resolution  and image acquiring procedure of these three techniques.

\subsection{Photoemission studies}

\begin{figure}[!ht]
 \begin{center}
\includegraphics[scale=1.8]{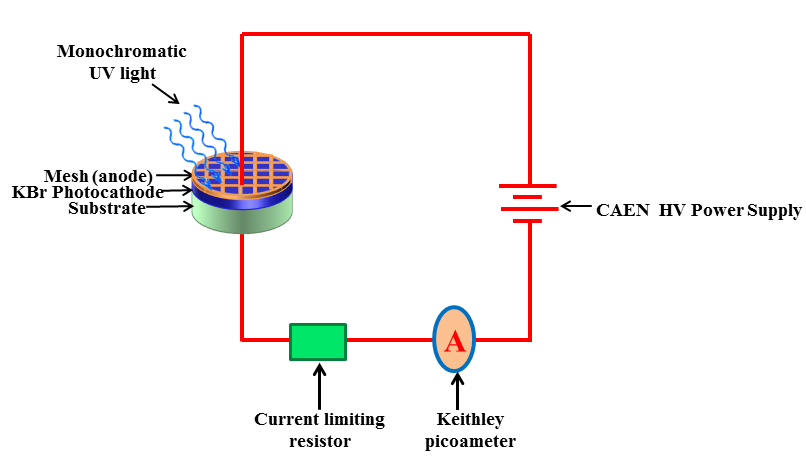}
 \caption{Circuit diagram of photoelectrons collection mode.}
     \label{XRD}
    \end{center}
  \end{figure}

To perform the photoemission studies, KBr films were deposited on stainless steel substrate. A representative circuit diagram of electron collection mode is shown in Figure 5. Ultraviolet light is illuminated on film surface and photoelectrons are generated . These ejected electrons are collected by an mesh anode which is kept at 3 mm distance and at positive bias voltage  with respect to the cathode. The resulting photoelectrons current was read by a Keithley picoammeter. A current limiting resistor was used to avoid the damage of picoammeter due to current fluctuation. 
  
The photocurrent of ``as deposited" KBr photocathode is measured in 120 to 180 nm spectral range. The spectral responses of all films are normalized with the response of 500 nm thick film and obtained results are shown in Figure 6 as a function of wavelengths. It is observed that the photocurrent has been increased with increasing film thickness while applied bias voltage and other external parameter are kept fixed. This increment in photocurrent is related to the ejection of more photoelectron from photocathode surface with increasing thickness. From X-ray diffraction analysis of these films, which we have reported in our earlier article~\cite{RRai}, it was found that defect density as well as microstrain decrease and therefore, film crystallinity improves with increasing thickness. The film with a better crystal structure and fewer impurities, ejected electrons will have a higher probability of survival through transport towards vacuum interface and contribute to the photocurrent. This may be one of the possible reasons behind the higher value of photocurrent for thicker films. Moreover this, from morphological measurements, it is observed that this variation in photocurrent is also related to the surface quality of films. The surface area coverage and grain density of 50 nm film is less in comparison with the thicker films and therefore it has the lowest value of photocurrent. When film thickness is increased more layers are grown on the substrate which results in an increment of the KBr coverage area and grain density. This enhances the numbers  of ejected electron and therefore the photocurrent value .

\begin{figure}[!ht]
 \begin{center}
\includegraphics[trim=0cm 0.0cm 0cm 0.0cm, clip=true, totalheight=0.34\textheight, angle=0]{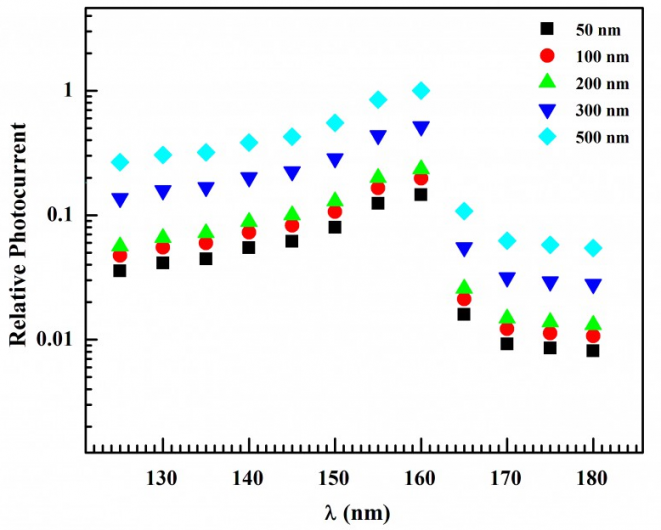} 
 \caption{Relative Photocurrent as a function of wavelengths with varying film thickness.}
     \label{XRD}
    \end{center}
  \end{figure}

 \section{Conclusion} 
The morphological and photoemissive behavior of KBr photocathodes was investigated in order to provide a clear understating regarding the thermal evaporation film fabrication process. The AFM analysis reveals that KBr thin films photocathode have granular structure and roughness of these films are altered with deposition of more layers. From SEM images, it is determined that grain density and surface area coverage are increased with film thickness. But increment in a grain size is observed only up to a certain thickness after that it gets saturated and this observation is also supported by TEM results. The electron diffraction patterns of KBr films reflect its crystalline nature. All the observed spectral rings can be indexed as those of KBr fcc crystallite. From a photoemission measurements, it is evident that the photocurrent increases with increasing film thickness. Moreover, from SEM, TEM and AFM results it is observed that this variation in photocurrent is also related to the surface morphology of  films. With an increment in the film thickness, the surface area coverage and grain density, photocurrent has been enhanced.

 \section{Acknowledgment}
  This work was partially supported by the Department of Science and Technology (DST) FIST, PURSE and the Indian Space Research Organization (ISRO) under ISRO-SSPS programs. R. Rai acknowledges University Grant Commission (UGC) and SERB-DST New Delhi, India for providing financial support.

\end{document}